# Measurements of Correlated Insulator Gaps in a Transition Metal Dichalcogenide Moiré Superlattice

Xiong Huang[1,2,†#], Dongxue Chen[3#], Zhen Lian[3#], Qiran Wu[1], Mina Rashetnia[1], Mark Blei[4], Takashi Taniguchi[5], Kenji Watanabe[6], Sefaattin Tongay[4], Su-Fei Shi[3,7*], Yong-Tao Cui[1*]

1. Department of Physics and Astronomy, University of California, Riverside, California, 92521, USA
2. Department of Materials Science and Engineering, University of California, Riverside, California, 92521, USA
3. Department of Chemical and Biological Engineering, Rensselaer Polytechnic Institute, Troy, NY 12180, USA
4. School for Engineering of Matter, Transport and Energy, Arizona State University, Tempe, AZ 85287, USA
5. Research Center for Materials Nanoarchitectonics, National Institute for Materials Science, 1-1 Namiki, Tsukuba 305-0044, Japan
6. Research Center for Electronic and Optical Materials, National Institute for Materials Science, 1-1 Namiki, Tsukuba 305-0044, Japan
7. Department of Electrical, Computer & Systems Engineering, Rensselaer Polytechnic Institute, Troy, NY 12180, USA

† Current Address: Department of Physics, Columbia University, New York, NY 10027, USA.

**Moiré superlattices of transitional metal dichalcogenides exhibit strong electron-electron interaction that has led to experimental observations of Mott insulators and generalized Wigner crystals. In this letter, we report direct measurements of the thermodynamic gaps of these correlated insulating states in a dual-gate $WS_2/WSe_2$ moiré bilayer. We employ the microwave impedance microscopy to probe the electronic features in both the graphene top gate and the moiré bilayer, from which we extract the doping dependence of the chemical potential of the moiré bilayer and the energy gaps for various correlated insulating states utilizing the Landau quantization of graphene. These gaps are relatively insensitive to the application of an external electric field to the $WS_2/WSe_2$ moiré bilayer.**

Moiré superlattices formed by two-dimensional (2D) transition metal dichalcogenide (TMD) layers have recently emerged as a promising platform to study the effects of strong correlation [1–4]. A plethora of novel electronic states has been discovered, including Mott insulators [5–8], generalized Wigner crystal insulators at fractional fillings [5,9–18], and dipolar excitonic insulators [19–23]. The strong electron correlation in these systems arises due to the formation of electronic flat bands in which the kinetic energy is substantially reduced and yields to the Coulomb interaction. Therefore, strong Coulomb repulsion can effectively localize carriers in the periodic moiré superlattice, leading to various charge-order states. Such correlated states can exhibit relatively high transition temperatures reflecting the strong correlation strength. For example, the Mott insulator state in the moiré superlattice of $WS_2/WSe_2$ survives above 180 K [11]. Although the temperature dependence study gives a rough estimate of the bandgap, however, it does not directly lead to the quantitative measurement of energy gaps of these insulating states.

To gain deeper insights into the understanding and the engineering of correlated states, it is beneficial to characterize the correlation strength in a quantitative way by measuring the energy gaps of the various correlated insulating states in TMD moiré systems [24–26]. Gap measurements in 2D correlated systems are in general challenging because the formation of the correlated states requires doping the system at appropriate carrier densities; hence, single particle spectroscopy techniques such as scanning tunneling spectroscopy and angle-resolved photoemission cannot be directly applied [27–29]. In this work, we report a direct measurement of thermodynamic gaps of the correlated insulating states in the $WS_2/WSe_2$ moiré superlattice by sensing the chemical potential of the moiré bilayer as a function of carrier doping [30–34]. The energy gap at the Mott insulator state at one hole per moiré unit cell is extracted to be 65 meV, and the gaps at fractional fillings of -1/3 and -2/3 are 10 meV. Away from the correlated insulating states, the system generally exhibits negative compressibility, a signature of strong electronic interaction.

The schematic of the device is shown in Fig. 1a. (See Supplementary Materials for device fabrication details [41]). A TMD heterobilayer is encapsulated by thin hBN flakes to form a dual-gate device geometry with monolayer graphene (MLG) as the top gate and few-layer graphite as the back gate. We perform microwave impedance microscopy (MIM) measurements by parking a sharp metal tip over the device [35–37]. A small microwave excitation (with a power of 1-10 uW and frequency of ~10 GHz) is applied to the tip via an impedance-matching network. Oscillating electric fields at microwave frequency are generated near the tip apex and screened by the sample underneath. Changes in the sample's local electrical properties, including dielectric constant and conductivity, will induce changes in the reflected microwave signals, of which the in-phase and out-of-phase components are acquired as the MIM-Re and MIM-Im signals, respectively. With its sensitivity to local conductivity, MIM has been utilized to study the correlated insulating phases in semiconducting moiré superlattices by tuning the carrier density with a single back gate [8,11]. In our dual-gate device, we find that when the MLG top gate is driven to the quantum Hall regime under a large out-of-plane magnetic field, its bulk conductivity drops such that the microwave electric fields from the tip cannot be completely screened by the MLG top gate and are thus able to reach the TMD heterobilayer and probe its conductivity [20]. The hBN thicknesses for the top and back gates are 5 and 16.3 nm, respectively, measured first by atomic force microscopy and later verified by the calibration of electron density via dual gate doping.

We record the MIM signals as a function of both top and bottom gate voltages to obtain a dual-gate map. A typical dual gate map of MIM-Im data measured at 6 T is presented in Fig. 1c. (See Fig. S1 of Supplementary Materials for more data at different magnetic fields [41]). There are two main sets of features. The first set of features is the diagonal lines corresponding to insulating states in the TMD moiré bilayer. In a dual-gate structure, the carrier density of the moiré bilayer depends on both top and bottom gate voltages, which can be described by the following equation:

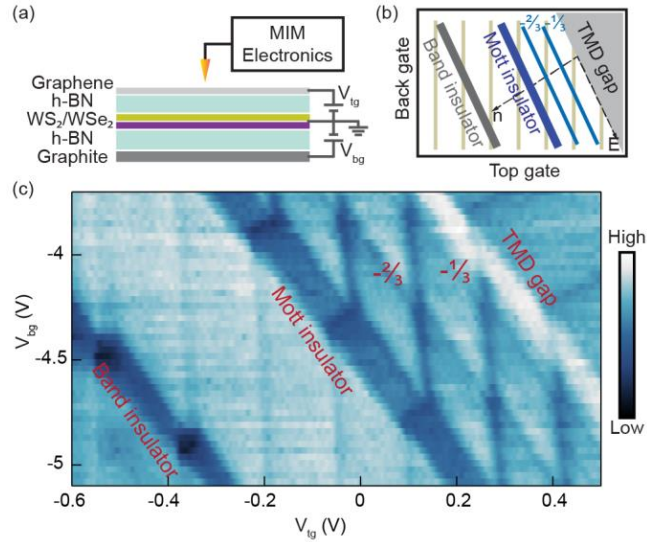

**Figure 1. Probing the correlated insulating states in a dual-gate WS$_2$/WSe$_2$ moire device with microwave impedance microscopy.** (a) Schematics of the measurement setup. (b) Expected features in a data map as a function of both top and back gate voltages. (c) Experimental data taken at 6 K and 6 T.

$$n_{moire} = C_{tg}(V_{tg} + \frac{\mu_G - \mu_{moire}}{e}) + C_{bg}(V_{bg} - \frac{\mu_{moire}}{e}) \quad (1)$$

Here, $C_{tg}$ ($C_{bg}$) is the geometric capacitance between the top gate (back gate) and the sample, $\mu_{moire}$ and $\mu_G$ the chemical potentials of the moiré bilayer and MLG, respectively, $e$ the electron charge, and $V_{tg}$ ($V_{bg}$) the voltage applied on the top (back) gate. The chemical potential change in the few-layer graphite back gate is minimal and has been neglected due to its large density of states. Each of these lines corresponds to a constant carrier density in the TMD moiré bilayer, and their pattern matches the expected filling factors, n, defined as the number of carriers per moiré unit cell, for the correlated insulating states in the WS$_2$/WSe$_2$ moire superlattice. The $n = -1$ state corresponds to a Mott insulator, the $n = -1/3$ and $-2/3$ states correspond to generalized Wigner crystal states, and the $n = -2$ state corresponds to the band insulator state with the Fermi level between the first and second moiré minibands in the Wse$_2$ layer.

The second set of features are a series of lines that appear at approximately constant top gate voltages which correspond to LL gaps in the MLG top gate [38]. The carrier density in the MLG graphene depends on the potential difference between the MLG and the TMD heterobilayer, which can be described by Equation 2

$$n_G = C_{tg}(\frac{\mu_{moire} - \mu_G}{e} - V_{tg}) \quad (2)$$

These LL features are approximately equally spaced with a carrier density of 5.8 x10$^{11}$ cm$^{-2}$ for graphene at the measurement magnetic field of 6 T. An illustration of these two sets of features is plotted in Fig. 1b. Based on Eq. (2), when the MLG is kept in a specific LL gap, both $n_G$ and $\mu_G$ are fixed, and we find a relation: $\Delta\mu_{moire} = e\Delta V_{tg}$. We thus can extract the $V_{tg}$ value for a particular LL as a function of $V_{bg}$, which are then converted to $\Delta\mu_{moire}$ vs $n_{moire}$. Figure 2(a) plots a high-resolution dual-gate MIM-Im map taken at 4 K and 6 T. To track the LLs, we take the derivative $d(MIM\text{-}Re)/dV_{tg}$ (Fig. 2b) in which the positions of LLs are sharply defined (See Fig. S2 of Supplementary Materials for raw MIM-Re data [41]). For each LL, we follow the procedure described above to extract the curves of $\Delta\mu_{moire}$ vs $n_{moire}$. Of particular interest are the crossings of LLs with the insulating states in the TMD moiré bilayer. We find that the LLs shift their positions when they cross the insulating states in the TMD moiré bilayer. This is because the chemical potential of the TMD moiré bilayer will change across an insulating gap, therefore, the top gate voltage must also change by the same amount to compensate for it so that the carrier density in MLG is maintained at the same value for each specific LL. This behavior of the graphene LL and the ability for MIM to track features in both MLG and the TMD moiré bilayer allow us to directly extract the energy gaps of the correlated insulating states. Figure 2c plots the processed $\Delta\mu_{moire}$ vs $n_{moire}$ curves extracted by tracking several LLs around the crossings. The abrupt jumps in these curves correspond to the energy gaps at various insulating states including the Mott insulating state at $-1$, the generalized Wigner crystals at $-1/3$, $-2/3$, and $-4/3$, as well as the band insulator gap at $-2$. The values of these gaps are listed in Table 1 together with the transition temperatures determined in back-gate only devices. Their general trends match qualitatively, but the quantitative ratio of $\Delta/k_B T_C$ varies, potentially due to the different natures of these insulating states. We also observe the $-4/3$ state in the derivative data in Fig. 2b but cannot resolve the gap value from the chemical potential curve, which suggests that the $-4/3$ gap is less than our measurement resolution of 2 meV.

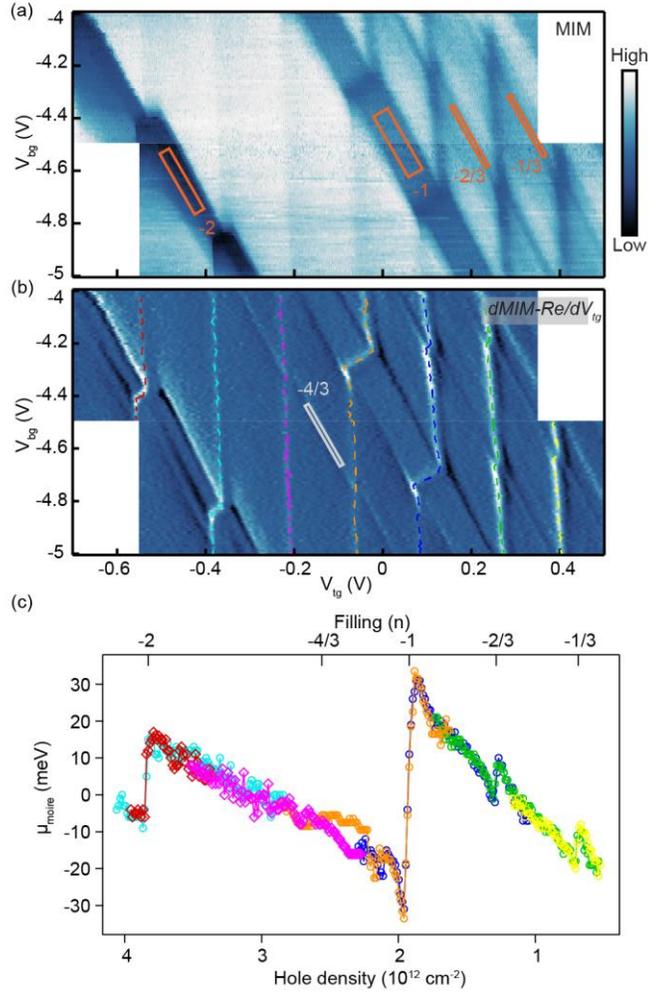

**Figure 2. Extraction of chemical potential and correlated insulator gaps in the TMD moiré bilayer.** (a) MIM-Im and (b) Derivative of MIM-Re, d(MIM-Re)/dVtg, as a function of both top and back gate voltages, measured at 4 K and 6 T. Dotted lines that track the LLs in the MLG are overlaid in (b). (c) Extracted chemical potentials of the TMD moiré bilayer plotted as a function of carrier density and the filling factor. Line colors correspond to specific LLs as marked in (b). The chemical potentials are offset by matching the average values in ranges with density overlap.

| $\nu$ | -1/3 | -2/3 | -1 | -4/3 | -2 |
|---|---|---|---|---|---|
| $\Delta\mu$ (meV) | 9.4±1.5 | 9.8±1.5 | 64.8±2.0 | <2.0 | 22.3±2.0 |
| $T_C$ (K) | 35 | 35 | 160 | - | 60 |

Table 1. Results of thermodynamic energy gaps of the correlated insulating states in the $WS_2/WSe_2$ moiré superlattice, compared with their transition temperatures [11].

Each crossing of a LL with an insulating state in the TMD bilayer will provide a measurement of the insulating gap. In the experimentally accessible gate range, the insulating states cross multiple LLs as seen in Fig. 1, and the extracted chemical potential profiles along these LLs are plotted together in Fig. 2c. These crossings correspond to different out-of-plane displacement electric fields on the TMD bilayer, and our measurement results do not show any appreciable changes in the gap size within the accessible range of displacement fields in our experiment.

Away from the insulator gaps, we find that the chemical potential generally decreases with increasing carrier density, exhibiting negative compressibility, which is particularly pronounced right before and after the insulating gap (as shown in Fig. 2c). This behavior is not expected from a single-particle picture in which adding carriers to an energy band should only increase the chemical potential, but a negative compressibility can occur in systems with strong electron-electron interactions [39,40]. The inverse electron compressibility, $\frac{\partial \mu}{\partial n}$, corresponds to the slope of the $\Delta\mu_{moire}$ vs $n_{moire}$ curve. Between $n=0$ and $n=-1$, its value is $-3.38 \times 10^{-11}\ meV \cdot cm^2$, while it is $-1.89 \times 10^{-11}\ meV \cdot cm^2$ in the filling range of $-2 < n < -1$. The larger, more negative value of $\frac{\partial \mu}{\partial n}$ for $-1 < n < 0$ than that for $-2 < n < -1$ indicates a stronger correlation strength for $-1 < n < 0$, which is consistent with the larger gap values for $n = -1/3$ and $-2/3$ than $n = -4/3$.

In summary, our measurements provide quantitative results on the thermodynamic energy gaps of the correlated states in the archetypal $WS_2/WSe_2$ moiré superlattice. The gap values for the $n = -1$ and $-2$ states are comparable to that in $WS_2/MoSe_2$ superlattice obtained through capacitance measurements [17], while the values for $n = -1/3, -2/3$, and $-4/3$ have not been reported previously. The MLG in our device geometry serves as both the sensor for chemical potentials and the top gate that can be used to tune carrier doping and electric field. This device structure is fully compatible with fabrication of other van der Waals devices, and we believe this technique can be readily applied to other similar 2D systems.

X.H., Q.W., and Y.-T.C. acknowledge support from NSF under award DMR-2104805 and DMR-2145735. Z. Lian and S.-F.S. acknowledge support from NYSTAR through Focus Center-NY–RPI Contract C150117. The device fabrication was supported by the Micro and Nanofabrication Clean Room (MNCR) at Rensselaer Polytechnic Institute (RPI). S.-F. S. also acknowledges the support from NSF Grant DMR-1945420, DMR-2104902, and ECCS-2139692. The optical spectroscopy measurements were supported by a DURIP award through Grant FA9550-20-1-0179. S.T. acknowledges support from NSF DMR-1904716, DMR-1838443, CMMI-1933214, and DOE-SC0020653. K.W. and T.T. acknowledge support from the JSPS KAKENHI (Grant Numbers 20H00354, 21H05233 and 23H02052) and World Premier International Research Center Initiative (WPI), MEXT, Japan.

[#] These authors contributed equally to this work.
[*] Corresponding authors: shis2@rpi.edu, yongtao.cui@ucr.edu.

**References**

[1]	F. Wu, T. Lovorn, E. Tutuc, and A. H. Macdonald, Phys Rev Lett **121**, 026402 (2018).

[2]	M. H. Naik and M. Jain, Phys Rev Lett **121**, 266401 (2018).

[3]	Y. Zhang, T. Liu, and L. Fu, Phys Rev B **103**, 155142 (2021).

[4]	B. Padhi, R. Chitra, and P. W. Phillips, Phys Rev B **103**, 125146 (2021).

[5]	E. C. Regan et al., Nature **579**, 359-363 (2020).

[6]	Y. Tang et al., Nature **579**, 353-358 (2020).

[7]	L. Wang et al., Nat Mater **19**, 861-866 (2020).

[8]	Z. Chu et al., Phys Rev Lett **125**, 186803 (2020).

[9]	Y. Xu, et al., Nature **587**, 214-218 (2020).

[10]	C. Jin et al., Nat Mater **20**, 940-944 (2021).

[11]	X. Huang et al., Nat Phys **17**, 715-719 (2021).

[12]	S. Miao et al., Nat Commun **12**, 3608 (2021).

[13]	Y. Shimazaki, et al., Nature **580**, 472-477 (2020).

[14]	A. Ghiotto et al., Nature **597**, 345-349 (2021).

[15]	T. Li et al., Nature **600**, 641-646 (2021).

[16]   E. Liu, et al., Phys Rev Lett **127**, 037402 (2021).

[17]   T. Li, et al., Nat Nanotechnol **16**, 1068 –1072 (2021).

[18]   H. Li et al., Nature **597**, 650 –654 (2021).

[19]   J. Gu, et al., Nat Phys **18**, 395 –400 (2022).

[20]   D. Chen et al., Nat Phys **18**, 1171 –1176 (2022).

[21]   Z. Zhang et al., Nat Phys **18**, 1214 –1220 (2022).

[22]   Y. Zeng, et al., Nat Mater **22**, 175 –179 (2023).

[23]   Q. Tan, et al., Nat Mater **22**, 605–611 (2023).

[24]   T. Li, et al., Nat Nanotechnol **16**, 1068 –1072 (2021).

[25]   H. Li et al., arXiv, arXiv:2209.12830

[26]   Z. Xia, et al., arXiv, arXiv:2304.09514

[27]   S. Xie et al., Sci Adv **8**, eabk1911(2022).

[28]   S. Ulstrup et al., Sci Adv **6**, eaay6104 (2020).

[29]   C. H. Stansbury et al., Sci Adv **7**, eabf4387 (2021).

[30]   K. Lee, et al., Science **345**,58-61(2014).

[31]   F. Yang, et al., Phys Rev Lett **126**, 156802 (2021).

[32]   B. Sun et al., et al., Nat Phys **18**, 94-99 (2022).

[33]   K. Yasuda, et al., Science **372**,1458-1462 (2021).

[34]   X. Wang, et al., Nat Nanotechnol **17**, 367-371 (2022).

[35]   Y. T. Cui, et al., Review of Scientific Instruments **87**, 063711 (2016).

[36]   Z. Chu, et al., Microwave Microscopy and Its Applications, Annu. Rev. Mater. Res. 50, 105 (2020).

[37]   M. E. Barber, et al., Nat Rev Phys 4, 61–74 (2022).

[38]   Y. T. Cui et al., Phys Rev Lett **117**, 186601 (2016).

[39]   J. P. Eisenstein, et al., Phys Rev Lett **68**, 674 (1992).

[40]   J. P. Eisenstein, et al., Phys Rev B **50**, 1760 (1994).

[41]   See Supplementary Materials [URL will be inserted by publisher] for details on device fabrication, calculation of carrier density, dual-gate MIM data at different magnetic field, raw data for Figure 2(b).